\documentclass[10pt, conference,compsocconf]{IEEEtran}
\IEEEoverridecommandlockouts
\pdfoutput=1

\usepackage[hyphens]{url}
\usepackage{cite}
\usepackage{balance}
\usepackage{amsmath,amssymb,amsfonts}
\usepackage{algorithmic}
\usepackage{graphicx}
\usepackage{textcomp}
\usepackage{xcolor}
\def\BibTeX{{\rm B\kern-.05em{\sc i\kern-.025em b}\kern-.08em
    T\kern-.1667em\lower.7ex\hbox{E}\kern-.125emX}}
\usepackage{soul}

\begin{document}

\title{Metaverse Security and Privacy: An Overview}

\author{Zefeng Chen$ ^{1,2} $, Jiayang Wu$ ^{3}$, Wensheng Gan$ ^{1,2*}$\thanks{\IEEEauthorrefmark{1}Corresponding author.},  Zhenlian Qi$ ^{4}$ \\ 
	\\
	$ ^{1} $Jinan University, Guangzhou 510632, China\\
	$ ^{2} $Pazhou Lab, Guangzhou 510330, China\\
	$ ^{3} $Guangdong Ocean University, Zhanjiang 524088, China\\
    $ ^{4} $Guangdong Eco-Engineering Polytechnic, Guangzhou 510520, China\\
    
	Email: \{czf1027, csjywu1, wsgan001\}@gmail.com, qzlhit@foxmail.com
}

% Corresponding author: Wensheng Gan, wsgan001@gmail.com

\maketitle

\begin{abstract}
    Metaverse is a living space and cyberspace that realizes the process of virtualizing and digitizing the real world. It integrates a plethora of existing technologies with the goal of being able to map the real world, even beyond the real world. Metaverse has a bright future and is expected to have many applications in various scenarios. The support of the Metaverse is based on numerous related technologies becoming mature. Hence, there is no doubt that the security risks of the development of the Metaverse may be more prominent and more complex. We present some Metaverse-related technologies and some potential security and privacy issues in the Metaverse. We present current solutions for Metaverse security and privacy derived from these technologies. In addition, we also raise some unresolved questions about the potential Metaverse. To summarize, this survey provides an in-depth review of the security and privacy issues raised by key technologies in Metaverse applications. We hope that this survey will provide insightful research directions and prospects for the Metaverse's development, particularly in terms of security and privacy protection in the Metaverse.
\end{abstract}

\begin{IEEEkeywords}
    Metaverse, cyber, security, privacy, overview
\end{IEEEkeywords}

\IEEEpeerreviewmaketitle

\section{Introduction}  \label{sec:introduction}
% \footnote{https://www.facebook.com/}

The development of computer science has improved the quality of human life and provided convenience for humans' daily lives. As a possible future product of computer science, the Metaverse has been widely used in the scientific community \cite{dionisio20133d}. Metaverse is a post-reality space that can realize the combination of physical reality and digital virtuality. It has a high application value \cite{mystakidis2022Metaverse}. The Metaverse has a high degree of freedom, which allows users to build and transform in this space. The Metaverse concept evolved from the concepts of "Metaverse" and "avatar", which were mentioned in the foreign science fiction work "Avalanche" in 1992 \cite{ning2021survey}. From the 1970s to the 1990s, a large number of open-world multiplayer games appeared and affected a generation. In fact, the open world, which can be understood as the game itself, formed the early basis of the Metaverse. In 2003, there was a game called \textit{Second Life} \cite{rymaszewski2007second}, which was a bit of a liberation from the idea of the real world. In this game, people could live in a virtual world and adjust their identities by having their own \textit{Doppelgänger}. Players couldn't make it come true in the real world. On October 28, 2021, the US social media company \textit{Facebook} was renamed as \textit{Meta}. 2021 is also known as the first year of the Metaverse. Today, the Metaverse is still gradually evolving and enriching. The concept is gradually evolving with the way humans communicate, as well as the development and integration of various technologies.

As a global user-created game platform, Roblox\footnote{https://developer.roblox.com/} identifies eight elements of the Metaverse: identity, friends, immersion, low latency, diversity, anytime, anywhere, economic systems, and civilization \cite{han2021analysis}. In fact, these concepts are vague, which means that the Metaverse hasn't been completely defined yet. Despite all this, the Metaverse still has various core technologies to promote its realization and development \cite{lim2022realizing}. For example, the Metaverse \cite{sun2022big,sun2022metaverse1,lin2022metaverse} is inseparable from big data \cite{gan2017data,gan2019survey}, artificial intelligence (AI) \cite{huynh2022artificial}, interaction technology \cite{zhao2022Metaverse}, digital twins \cite{lv2022blocknet}, cloud computing \cite{seok2021analysis}, Internet of Things (IoT) \cite{mozumder2022overview}, blockchain \cite{gadekallu2022blockchain}, and so on. To be specific, it builds the economic system based on blockchain \cite{gadekallu2022blockchain}, provides an immersive experience based on interactive technology \cite{zhao2022Metaverse}, generates a mirror image of the real world based on digital twins \cite{lv2022blocknet}, realizes data computing, storage, processing and sharing based on cloud computing \cite{seok2021analysis}, and realizes interconnected intelligence based on AI \cite{huynh2022artificial} and IoT technology \cite{mozumder2022overview}. With such a grand vision, the Metaverse is bound to bring about many changes in human life. For example, in games, the Metaverse will provide an immersive gaming experience. In education, classrooms are everywhere and are no longer restricted by geographical factors. With the blessing of the Metaverse, travel enthusiasts can enjoy the fun of traveling without leaving their homes, and they are no longer restricted by the COVID-19 virus in the post-epidemic era. The Metaverse can make human lives better.

\begin{figure}[h]
	\centering
	\includegraphics[scale = 0.23]{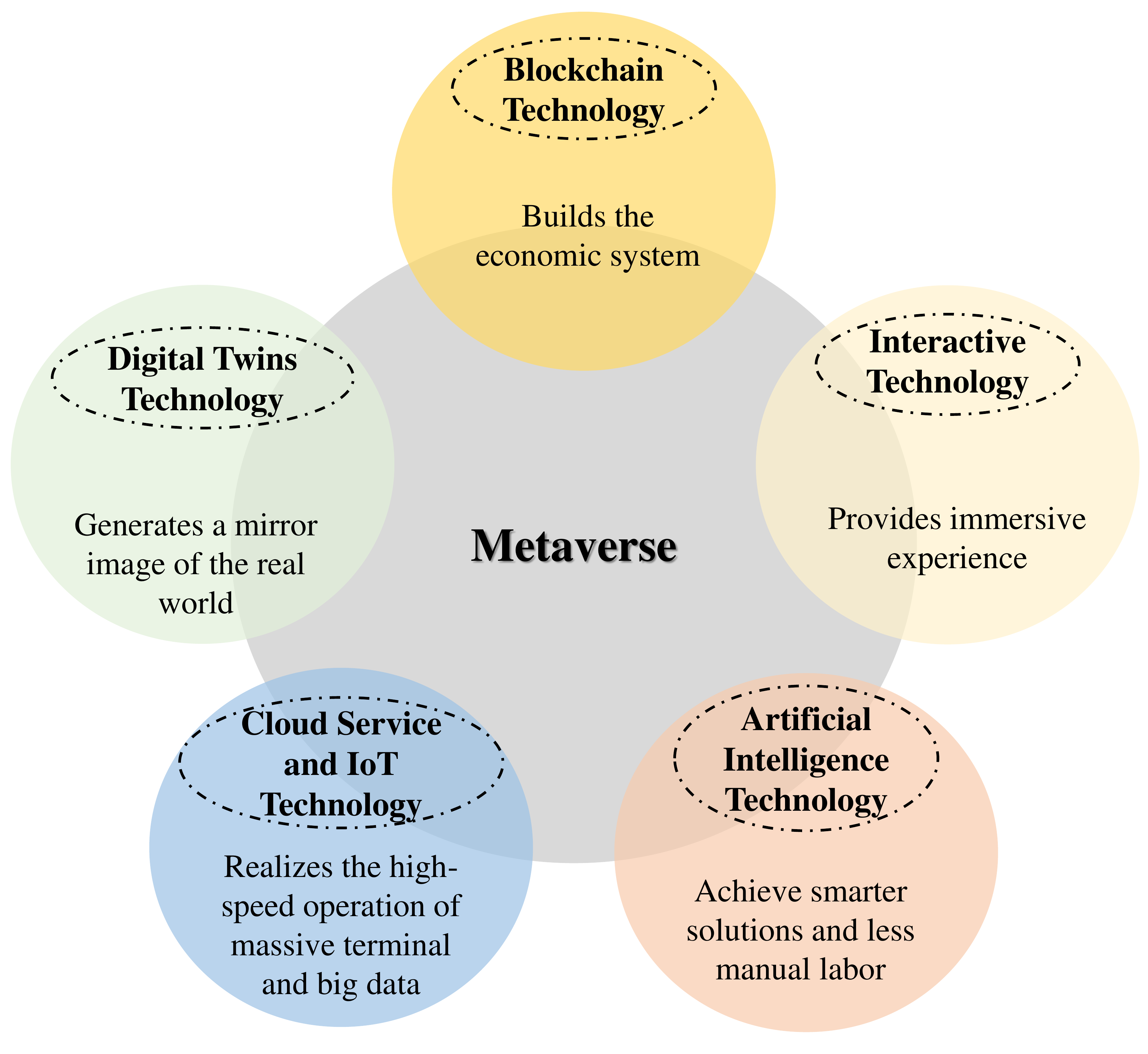}
	\caption{Metaverse-related technologies and their impact on the Metaverse.}
	\label{fig:Technology}
\end{figure}

Although the blueprint of the Metaverse is beautiful, with the rapid development of Metaverse-related technologies, it is foreseeable that more problems will follow \cite{kye2021educational}. The most important issues are security and privacy in the Metaverse, and they are about everyone's property and privacy protection in the future \cite{di2021Metaverse, wang2022survey}. The Metaverse Security Governance Shanghai Initiative proposes eight goals and principles and their basic requirements, including virtual and real progress, order protection, robust development, standardization of institutions, respect for privacy, safeguarding the future, rationality and pragmatism, and openness and collaboration. Existing Metaverse-related technologies have emerged to counter some of the issues of security and privacy in the Metaverse, but some remain unsolved.

In this paper, we first introduce the advantages and application scenarios of the Metaverse and propose the composition of the related technologies of the Metaverse. And then, we raise some potential security and privacy concerns in the Metaverse in detail. Besides, we further discuss the current solutions for the security and privacy of the Metaverse from the perspective of related technologies. Finally, we highlight some potential unresolved questions about the security and privacy issues raised by key technologies in Metaverse applications.

The rest of this paper is organized as follows. In Section \ref{sec:Metaverse}, we introduce the strengths, applications, and composition of related technologies in the Metaverse. In Section \ref{sec:security}, we describe the security and privacy problems of the Metaverse and how related technologies address these problems. In Section \ref{sec:opportunities}, based on the above sections, we raise and discuss possible unresolved security and privacy issues in the Metaverse. Finally, we provide our conclusions in Section \ref{sec:conclusions}.

\section{Metaverse} \label{sec:Metaverse}
\subsection{Advantages from Metaverse}

The Metaverse will play an important role in human life in the future because it offers a variety of novel experiences \cite{lee2021metaverse}. It has many advantages and opens up many new possibilities \cite{park2022Metaverse}, which we will describe below:

\subsubsection{Strong virtual identity} In the Metaverse, the sense of virtual identity substitution is unprecedentedly powerful. Furthermore, virtual activities in the Metaverse can help users develop a stronger sense of self \cite{lau2014gender}. In the Metaverse, the virtual identity is consistent throughout the space, and the customized avatar can produce a sense of uniqueness and immersive reality \cite{dincelli2022immersive}.

\subsubsection{Immersive experience} With augmented reality (AR)\footnote{https://en.wikipedia.org/wiki/Augmented\_reality} and virtual reality (VR)\footnote{https://en.wikipedia.org/wiki/Virtual\_reality}, users can have a sensory-immersive experience in the Metaverse \cite{dincelli2022immersive, kozinets2022immersive}. For example, when experiencing a fire, users can see the light of the fire in the sense of vision, smell the burning fire in the sense of smell, hear the sound in the auditory sense, and feel the temperature increase in the sense of touch.

\subsubsection{Bring a wider range of social activities} The Metaverse can provide rich online social scenes and ignore the distance between people. Based on a strong sense of identity and immersive experience, people's social activities are no longer limited to ordinary social activities and can widely try various novel activities \cite{ko2021study}. Moreover, friends who are far away from a foreign country can also participate in such social activities regardless of geographic distance.

\subsubsection{Virtual economic widely circulate} Today, most of our virtual currencies are circulated in games and cannot be withdrawn or consumed across platforms. The Metaverse has an economic system similar to the real world \cite{duan2021Metaverse}. The user's virtual property is more easily guaranteed than on the gaming platform and can be widely circulated without the constraints of the platform.

\subsubsection{Openness to free creation} The Metaverse is all-inclusive and includes various things. As a result, user creation and innovation are critical to the Metaverse's updating iteration \cite{ondrejka2004escaping}. The Metaverse opens up users' creation as the leading role. Thus, the content of the Metaverse becomes richer and more striking.

\subsection{Main Changes in Metaverse}

The future world has brought richer possibilities precisely because of the aforementioned benefits of the Metaverse. The coming Metaverse will undoubtedly bring great changes to human life and social and economic development. Just as 5G led to the development of unmanned driving technology, it is believed that the Metaverse will lead to new technology applications in various industries \cite{allam2022Metaverse}. The changes brought about by the Metaverse can be summarized as "three new", which are a series of new technologies, new forms, and new business models. The main changes are as follows:

\subsubsection{Technological innovation and cooperation} The emergence of the Metaverse can further improve the efficiency of social production from the perspective of technological innovation and cooperation \cite{kraus2022facebook}. For example, in the post-pandemic era, many production studies are at a standstill. However, if there were a space like Metaverse that could replicate the real world, then production research could be carried out in the Metaverse. Besides, from the perspective of technological cooperation, experts in the same research field from different companies and even different countries can also do research together in the Metaverse, which will greatly promote technological cooperation and innovation.

\subsubsection{Work environment} After the Metaverse has matured, the environment in which people work will be changed. Obviously, people will work in the virtual world instead of the real world. As the Metaverse matured, the virtual world could duplicate or even surpass the real world \cite{rospigliosi2022Metaverse}. Hence, a lot of work can be done in the virtual world. What's more, when a task in the virtual world goes wrong, the ability to return to the previous step, as in the game, reduces the cost of getting it wrong.
    
\subsubsection{Contribution to creative industry} The Metaverse can promote the derivation of the creative industry \cite{mystakidis2022metaverse}. For example, the Metaverse will undoubtedly boost the consumption of cultural products like games \cite{shin2022actualization}, comics \cite{thawonmas2011camerawork} and science fiction movies \cite{ahn2022bifold}. As the predecessors of the Metaverse, their development has promoted the development of the Metaverse to a certain extent. When the Metaverse develops, these kinds of products are bound to be very popular.
    
\subsubsection{Cultural tourism and information consumption} It is an important trend in the tourism industry to promote cultural tourism projects and greatly stimulate information consumption \cite{buhalis2022mixed}. Consumer behavior is driven by cultural tourism and related cultural industries. However, the cultural tourism industry is suppressed due to the raging epidemic. With the improvement of AR/VR equipment through interactive technology, travelers in the Metaverse can enjoy the joy of immersive travel. It will also promote information consumption to a certain extent.
    
\subsubsection{Promotion of smart cities} With the support of cloud computing and digital twins technology, the completion of the Metaverse will promote the construction of smart cities and innovate the social governance model. Smart cities connect facilities through technologies such as AI and IoT, aiming to provide better services to citizens and better manage cities \cite{song2021build}. The Metaverse can provide the following support for smart cities: global monitoring, simulated city governance decision-making, emergency event handling decision-making, and simulation of urban planning and construction \cite{allam2022Metaverse}.

\subsection{Application Scenario}

Note that digital services have brought great convenience to human life in the $21^{\rm st}$ century. The Metaverse hopes to use more mature and diversified technologies to further improve the quality of life.

\subsubsection{Handle official business} Recently, home offices have become the norm of lives. People can complete their tasks given by the company at home through the Internet. However, there are some obvious disadvantages, such as difficulty in collaboration and low efficiency of communication. Some manufacturers are committed to solving these problems. \textit{Microsoft} presented the project \textit{Mesh for Microsoft Teams} in which users can complete the tasks of the job in the virtual space by combining holographic projection, such as holding meetings, sending information, and processing shared documents. In addition, \textit{NVIDIA} presented the \textit{Omniverse} cloud service, through which designers across the world can complete real-time collaborative design without time or space constraints.

\subsubsection{Education} The combination of Metaverse and online education provide possibility to make up for the shortcomings of online and offline education. For young children, teachers can define their own classroom models according to the needs of the curriculum. It allows students to switch to different teaching environments without being limited by time or space. Higher education students and workers must prioritize the development of practical skills. In the case of Inwak Meta University, it was founded to teach employability training through a virtual immersion platform.

\subsubsection{Tourism} People like traveling, mostly because they like to learn about the cultural landscape of different cities. In the future, Metaverse tourism may bring travelers totally different experiences. For example, based on the AI response system, it can answer all kinds of questions about the scenic spots on the journey in the guidance service. In addition, people can reproduce the historical scenes, and travel enthusiasts can talk to the famous historical figures through Metaverse. Moreover, the interactive activities will be increased so that they can explore the original cave.

\subsubsection{Social intercourse} At present, the products of the Metaverse social intercourse concept are mainly used in the gaming field. For example, \textit{Second Life} was created in 2003. This game integrates plenty of details from real life into the game. The attraction of this game is that the players can start a second life in the virtual world as their avatars. The players can do some social activities, such as making friends and creating and trading goods.

\subsubsection{Shopping} There are two scenarios that are widely used in shopping. When buying clothes, with the VR equipment, the consumer can choose their favorite clothes in the appropriate sizes, which can be projected onto the body. Furthermore, they can change the background and observe the dressing effect in various spaces. When buying cars, the consumer can see the details of the cars from different directions, which is an important indicator of whether to go to physical stores.

\subsection{Technological Composition} 

Since 2021, the concept of the Metaverse has attracted more investors' attention. They invested a lot in the companies to develop the Metaverse products. Investors are full of confidence in building the Metaverse, which is inseparable from their trust in current technology. It is possible to create more Metaverse products by combining current technologies, including artificial intelligence (AI), blockchain, interaction, cloud services, the Internet of Things (IoT), and digital twins.

\subsubsection{Blockchain} Actually, the blockchain is a vital part of the Metaverse. The technical value of the blockchain will continue to promote the deep integration of the real economy and the digital economy in the Metaverse, providing an open and transparent collaboration mechanism for the Metaverse \cite{gadekallu2022blockchain}. In particular, the decentralized feature of blockchain makes value creation and value transfer in the Metaverse more free and efficient \cite{puthal2018blockchain}.

\subsubsection{Interaction} At present, the most popular interactive technologies are augmented reality (AR) technology and virtual reality (VR) technology \cite{ardiny2018role}. The task of AR technology is to combine virtual and real objects, realize real-time interaction, and superimpose computer-generated virtual objects into the real world \cite{azuma1997survey}. The task of VR technology is to create a virtual three-dimensional interactive scene, allowing users to experience the virtual world with the help of the device, so that the user is immersed in the virtual world created by the device without feeling violated \cite{schuemie2001research}. Because AR and VR interactive technology are the keys to achieving an immersive experience, they are crucial technologies of the Metaverse. At present, \textit{Facebook} (\textit{Meta})\footnote{https://www.facebook.com/} has started to build a comprehensive platform from VR hardware, \textit{Microsoft}\footnote{https://www.microsoft.com/} has also begun to focus on the AR field, relying on their products \textit{Hololens}\footnote{https://www.microsoft.com/hololens} to improve the commercial capability of AR. Apple\footnote{https://www.apple.com/} has also started to layout in various scenes of AR and VR based on various hardware and software. These illustrate that AR and VR interactive technologies have some applications in the preliminary concept of the Metaverse \cite{lee2021metaverse}.

\subsubsection{Cloud service and Internet of Things (IoT)} Cloud service is a process of decomposing huge data processing programs into small programs through the network cloud, and then processing them through a system composed of multiple servers \cite{qian2009cloud}. And the Metaverse will inevitably bring very large data throughput \cite{sun2022big}, which is a big challenge in terms of computing speed \cite{cai2022compute}. Hence, cloud service technology can bring users a better experience in the Metaverse: service-oriented low latency, more efficient cooperation, and products built on the cloud that are easily accessible to users from all over the world. The Internet of Things (IoT) aims to make all physical objects that can be addressed independently, i.e., the Internet of Everything, based on the Internet's information carrier and telecommunication network. In the era of the Metaverse, massive sensors embedded in all kinds of devices collect information all the time, accompanied by massive information generation \cite{guan2022extended}. Cloud computing provides the IoT with internet infrastructure and powerful working abilities. The combination with cloud computing is the inevitable trend of the development of the IoT, and this will be the great driving force of the Metaverse.

\subsubsection{Artificial intelligence (AI)} AI is one of the key technologies to construct the Metaverse \cite{huynh2022artificial}. The Metaverse itself exists in a virtual form, and AI technology can give it realistic conditions for its actual existence \cite{aggarwal2022has}. To be specific, AI technology can process a large amount of data generated by users' activities in the Metaverse, which is mainly manifested as: generating AI models and creating virtual environments; mapping body movements to make virtual and real interaction more natural; synchronous translation of voice; and enhancing user interaction and participation \cite{lee2021all}. Furthermore, Metaverse will appear as a digital extension of the underlying technology in the process of practical application of the new need to solve the problems and requirements, which makes the AI technology match with its actual application demand, leading to an update of the progress of the AI technology innovation and causing it to develop new application fields \cite{haenlein2019brief}.

\subsubsection{Digital twins} It is a concept that refers to the construction of a digital model of a physical object before it has been built \cite{jiang2021industrial}. Users can simulate in virtual space and transmit the actual parameters to the real world. The impact of digital twins on the Metaverse can be summarized in the following four aspects: Reality becomes computable, interpersonal communication becomes more flexible, organizations become more flexible, and processes become more optimized \cite{hartmann2021digital}.

\section{Metaverse Security and Privacy Technology}
\label{sec:security}

\begin{figure}[h]
	\centering
	\includegraphics[scale = 0.21]{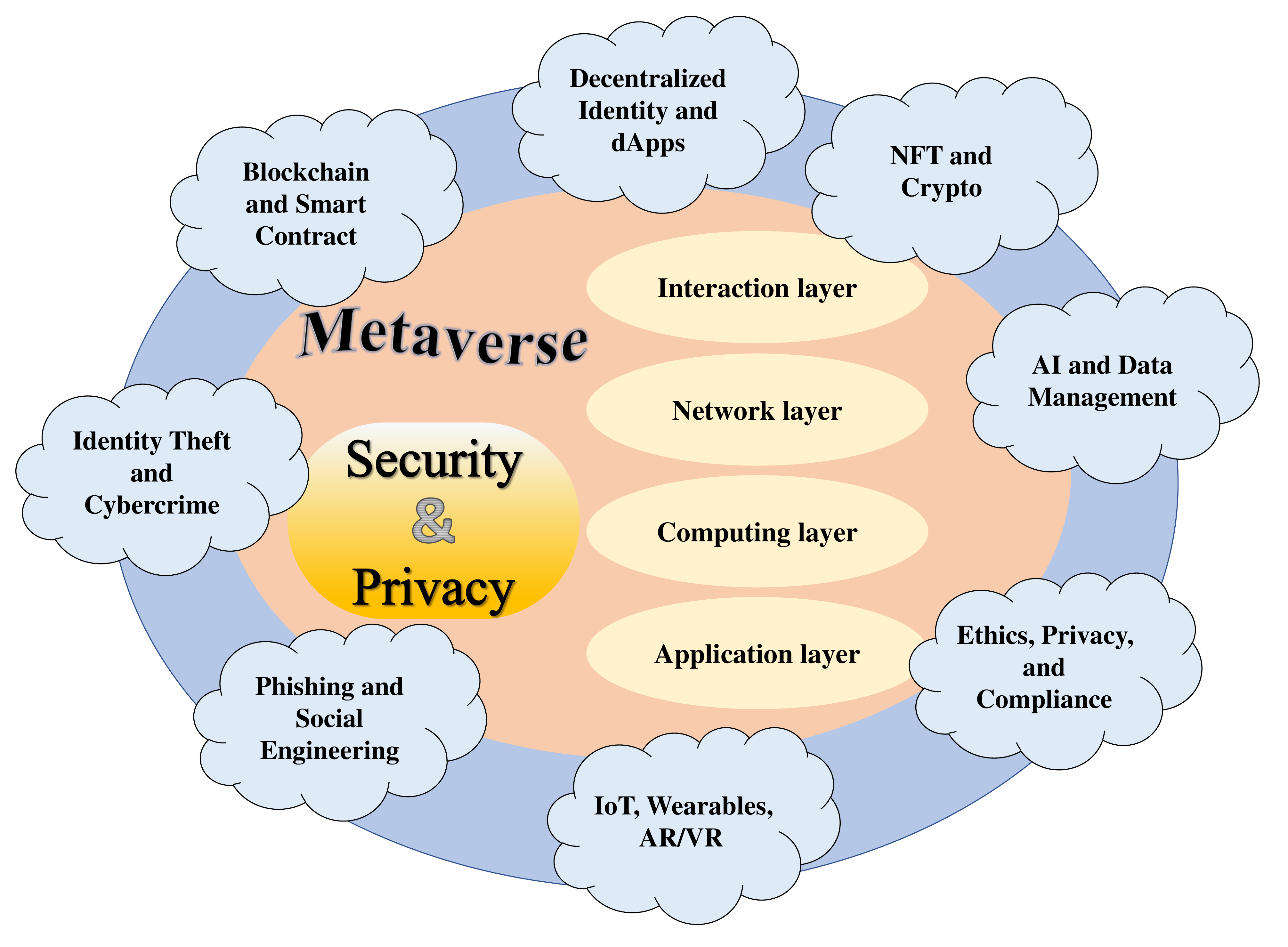}
	\caption{Four layers and eight major threats in virtual world of Metaverse.}
	\label{fig:Security}
\end{figure}

The Metaverse will profoundly affect the various needs of human beings. Thus, the security and privacy issues of the Metaverse cannot be ignored. As Henry Bagdasarian, Founder of Identity Management Institute\footnote{https://identitymanagementinstitute.org/}, once said, ``Metaverse will completely change the way people live, socialize, and conduct business, thus presenting new security challenges in the evolving digital world." As the new digital space takes shape to fully transition our physical world into the digital realm, experts in the field will be concerned with emerging Metaverse security risks, which will include new forms of security and data protection threats, identity theft, and fraud. Actually, security and privacy in the virtual world of Metaverse mainly occur in four layers. According to the current technological composition of the Metaverse, the security risks and threats of the Metaverse mainly include eight points, which are shown in Fig.\ref{fig:Security}. The following sections discuss the Metaverse security and privacy from the standpoint of the five main related technologies.

\subsection{Blockchain Security}

The leading Internet companies, such as Facebook and Google, hold a large amount of user data. Facebook has lost nearly \$50 billion in market cap since the data scandal\footnote{https://www.vox.com/2018/3/20/17144130/facebook-stock-wall-street-billion-market-cap}. Google prioritizes blockchain research on cloud storage services, while Facebook prioritizes secure payment. Blockchain adopts point-to-point transmission, consensus mechanism, and encryption algorithm technology to provide users with safe and reliable data storage and data transmission \cite{mingxiao2017review}. Hackers can still attack and destroy the system due to its insufficient design, such as the consensus mechanism, smart contract, or cryptographic algorithm.

\subsubsection{Consensus mechanism} The blockchain is a decentralized ledger, and the consensus mechanism helps to achieve consistency and correctness of data on different nodes. The popular consensus mechanism algorithms include PBFT, PoW, PoS, and DPoS \cite{mingxiao2017review}. However, there are security risks in these algorithms. For the PoW, it may suffer from the double spend attack\footnote{https://en.wikipedia.org/wiki/Double\-spending} that the attacker can control the entire blockchain if the attacker's computing power exceeds 51\% of the entire blockchain.

\subsubsection{Smart contract} Smart contracts are used to help users control their own assets and receive assets from the outside \cite{zheng2020overview}. The contract is open and transparent, which helps to improve transparency and prevent fraud. However, not everyone wants the confidential contract information disclosed. In addition, on the blockchain, developers can't modify the bugs in the protocol because it becomes immutable when deployed into the blockchain. Wormhole, which connects the two blockchains of Ethereum and Solana, was attacked by hackers, and its loss is estimated to be at least 320 million dollars\footnote{https://www.cnbc.com/2022/02/02/320\-million\-stolen\-from\-wormhole\-bridge\-linking\-solana\-and\-ethereum.html}.

\subsubsection{Cryptographic algorithm} Encryption ensures that the blockchain can't be tampered with. To generate digital signatures for secure transactions, blockchain primarily employs elliptic curve encryption algorithms (such as ECDSA, RSA, and DSA) \cite{zhao2019practical}. With the development of quantum computing, these algorithms can not guarantee security, so more and more researchers are beginning to pay attention to how to resist quantum attacks \cite{dinh2011mceliece}.

\subsection{Interactive Technology Security}  AR and VR with interactive technology are important parts of the Metaverse vision \cite{kozinets2022immersive}. However, it also has some security issues. There are two main challenges in interactive technology. One is that the amount of information in interactive technology is larger, so it is more likely to be stolen by criminals, and the other is that identity authentication is more difficult and insecure in the virtual world.

\subsubsection{Information security}  In recent years, face recognition technology has been widely used in identity authentication. However, there are still loopholes in facial recognition technology. For example, hackers can download photos from social accounts and  utilize these photos to create VR models of people's faces, thereby defeating the security systems of facial recognition technology. As AR/VR interactive technology is widely used in the Metaverse, more personal data will inevitably be generated, and there will also be more data interfaces. AR collects a lot of information about who users are and what users are doing, which is deeper than social media networks or other forms of technology. It has raised concerns and questions from users. If a hacker gains access to the device, the potential loss of privacy is beyond estimation.

As with AR, information security is a major concern for VR as well. A key concern in information security for VR is the highly personal nature of the data collected, such as biometric data like iris or retina scans, fingerprints, handprints, face geometry, and so on. For example, VR is able to do finger tracking and eye tracking. For example, when users use their fingers to type codes on the virtual keyboard, the VR system records and sends the finger and track data showing the fingers typing the password \cite{shah2014survey}. If an attacker could capture that data, they would be able to recreate the users' passwords. Some VR headsets may also feature eye tracking capabilities \cite{clay2019eye}. The data can provide additional value to malicious actors. If attackers can know exactly what a user is viewing, they can capture potentially valuable sources of information and steal it. Therefore, it is easy to foresee that guaranteeing personal data from being stolen by hackers will be a quite difficult task. For example, some sensors mounted on AR and VR devices can be used to take photos, which may allow hackers to record the users' movements in the Metaverse. It creates a certain potential security threat. For example, Chen \textit{et al.} \cite{chen2018case} proposed a 3D visual attack on digital passwords, which indicates that there are certain security risks with interactive devices. In essence, the security of information in interactive technologies still depends on the security of the data. Since 1975, data security was developed rapidly \cite{denning1982cryptography}. Until now, technologies related to data security have been widely developed, such as data security in cryptography \cite{denning1982cryptography, saleh2016data}, image and video data encryption method \cite{pommer2003selective, wajgade2013enhancing}, data security in cloud computing \cite{sood2012combined, thabit2021new}, and data security in cyberspace \cite{dobak2021thoughts, yang2022application}.

\subsubsection{Authorization} Nowadays, the authorization of AR and VR devices is usually ignored or performed on a smartphone or PC \cite{chan2015glass}. However, authenticating in a virtual space by leveraging traditional authentication techniques is not convenient. For example, if users need to perform identity authentication in an immersive meeting, they should take off the interactive product, which will greatly reduce their immersive experience. Many technologies that can be used for identity authentication have been developed, but there are still some security risks associated with identity fraud. For example, traditional methods like entering passwords are cumbersome and less suitable for virtual worlds. In addition, although the use of voice recognition technology can effectively authenticate an identity, it still has security risks with the development of voice synthesis technology. Therefore, biometric authentication using sensors is widely used. Among them, gait recognition methods \cite{gafurov2006biometric} designed for continuous authentication, thus improving safety. Besides, Duezguen \textit{et al.} \cite{duezguen2020towards} proposed a shoulder-surfing resistant authentication scheme that relies only on the equipment of the AR/VR head-mounted display, which improves security and authentication efficiency. Other study \cite{viswanathan2022security} suggested that authentication is no longer just the step of confirming identity; instead, multidimensional authentication is more applicable. These discoveries provide a certain technical basis for identity authentication, which ensures the security of the Metaverse. In addition, users need to be vigilant when using AR and VR devices, including:

\begin{itemize}
    \item Be careful to avoid disclosing private information that is too personal.
    \item Use a VPN to adopt advanced encryption and change IP addresses to keep your identity and data private.
    \item Keep firmware updated to fix security flaws in a timely manner and use comprehensive and trustworthy antivirus software.
    \item Pay attention to reviewing privacy policies.
\end{itemize}

\subsection{Cloud service and Internet of Things (IoT) security} Cloud services and IoT both have the ability to process large-scale data. However, due to their decentralized deployment, it is easy to have management problems, which could be attacked. Therefore, the security of infrastructure and data should be addressed.

\subsubsection{Infrastructure} Enterprises usually use remote monitoring mode to monitor the power consumption of servers \cite{wu2011energy}. The security level of these systems is usually low. If the cooling system was disrupted, then the servers may be experiencing downtime due to their excessive temperature\footnote{https://blog.cyble.com/2022/01/27/data\-centers\-facing\-risk\-of\-cyberattacks/}. Therefore, enterprises should improve the security level of their remote control management system and upgrade the hardware in due time. In addition, the infrastructure is prone to suffering large traffic attacks, such as DDoS\footnote{https://en.wikipedia.org/wiki/Denial\-of\-service\_attack}. It can build a reliable risk assessment platform to distinguish attack traffic from normal traffic. If it discovers the attack traffic, it could use Anycast \cite{moura2016anycast} to distribute attack traffic to the distributed server network.

\subsubsection{Data} Users can store their own data in the cloud and share it with others. Users with permission can access this data anytime and anywhere. In order to protect the data, protective measures should be deployed across all links, such as data storage and transmission. In data storage, the data should be encrypted so that even if hackers obtain the data, they cannot understand the contents \cite{rao2016study}. Moreover, when using zero-knowledge encryption \cite{yi2012single,adelsbach2001zero}, it can store users' encryption keys outside the cloud. Therefore, they cannot read the data even if the hackers get permission from the data server because they have no encryption key. After data encryption, the previous data retrieval technologies have failed. It can improve retrieval efficiency by using the encrypted information retrieval technology \cite{thandaiahprabu2022efficient}, such as linear search algorithm, security index, and keyword public key search. To ensure confidentiality and data integrity, data transmission must be encrypted using relevant security protocols such as PGP, SSL/TLS, IPSec, and others\footnote{https://msatechnosoft.in/blog/internet-security-ipsecurity-ssl-tls-pgp-firewall/}.

\subsection{Artificial Intelligence Security}

Nowadays, many AI algorithms can successfully complete the tasks of classification and recognition. However, because most algorithms have the black box characteristics, it becomes difficult to detect attacks.

\subsubsection{Generative adversarial network (GAN)} GAN\footnote{https://en.wikipedia.org/wiki/Generative\_adversarial\_network} is one of the most popular attacks that generates deceptive data through the iterative fusion of two neural networks \cite{yan2019method}. It can help solve this problem by strengthening the detection of malicious data. Ioannidis \textit{et al.} \cite{ioannidis2019graphsac} presented a way to recover the original data by detecting and eliminating malicious nodes and edges. Another method adds disturbance samples into the training step to improve the generalization ability of the model. Dai \textit{et al.} \cite{dai2019adversarial} presented an adversarial training method, adding the disturbance to the embedded space.

\subsubsection{Embedded malware} When hackers use the hiding technology EvilModel to embed fraud data into the training process, this is referred to as embedded malware \cite{wang2021evilmodel}. If the developers can't well understand the algorithms' principles, they will not discover the virus from the slight difference in the experimental results. At present, if the infected layer is not frozen, the virus can be directly destroyed by changing the parameter values and retraining the model. However, most developers may not change the pre-training parameters unless they want to make other applications. Therefore, it needs new ways to deal with these security threats. Recently, many studies \cite{li2021framework,luo2020adversarial,wang2017adversary} have focused on constructing a complete framework that can help developers find weaknesses in the machine learning pipeline and repair security vulnerabilities. Furthermore, an increasing number of studies are attempting to improve the interpretability of their modules \cite{kim2020transparency}.

\subsection{Digital Twins Security}

As stated above, security issues are widespread in the Metaverse. However, at the same time, the development of the Metaverse can also affect the security of systems in the real world. In fact, the Metaverse also represents a hypothetical parallel virtual world that can serve as a virtual form of the real world. The Metaverse can also generate some safe simulations of the real world, inspired by the digital twins \cite{tao2018digital}, which was originally proposed by NASA\footnote{https://www.nasa.gov/} in 1970. It is a digital technology that establishes a physical system within an information platform to simulate a physical entity, process, or system \cite{josifovska2019reference}. Actually, the Metaverse is closely related to digital twins conceptually in that the Metaverse is actually a digital twin of the entire real world \cite{nguyen2022toward}, but the Metaverse emphasizes the combination of virtual and real more. Therefore, the security of the digital twins is also an important topic that contributes to the security of the Metaverse. Current work has explored the security support of digital twins technology for cyber-physical systems (CPS) \cite{kim2021survey}, which can provide some reference for the security of the Metaverse. According to the life cycle of CPS, the security assurance of digital twins can be mainly divided into several stages \cite{eckhart2019digital}, which are respectively secure design of cyber-physical systems, intrusion detection, detecting hardware and software misconfigurations, security testing, privacy, system testing and training, secure decommissioning, and legal compliance. The technology of digital twins can be applied to the Metaverse to strengthen the close connection between the virtual and real worlds.

1) Since digital twins exist only virtually and often run in an environment isolated from real-time systems, they can also be used as testing and training platforms. Producers can test defenses before putting their products into use, or train them on how to deal with regular cyberattacks, which is helpful for risk assessments. Analyzing how the system to be engineered behaves under attack would allow engineers to estimate potential damage and dangers. Bécue \textit{et al.} \cite{becue2018cyberfactory} proposed the method of using digital twins in combination with a cyber range. In addition, another viable technical use case is to assess how to limit damage to the system, such as by simulating attack scenarios to prepare a containment strategy for compromised devices. Eckhart \textit{et al.} \cite{eckhart2018specification} demonstrate how the concept of digital twins can be leveraged to construct intrusion detection systems. An intrusion can be detected simply by comparing the inputs and outputs of the physical device with the digital twins. In the Metaverse, real-world threats can be simulated. The Metaverse can simulate and anticipate real-world threats through modeling to create strategies and detect intrusions by comparing input and output to deal with potential risks.
    
2) The hardware and software of the device are simulated in the digital twins, and these virtual representations can simulate the device's functions to a certain extent. By applying the same method to the configuration data, engineers can check whether the software is correctly configured \cite{eckhart2018towards}. In the future, with the gradual maturity of virtual reality and digitalization, there will be more security problems in all kinds of software and hardware. To utilize this detection of digital twins, a wide variety of hardware and software can be detected in the Metaverse in a similar way.
    
3) Building a security test platform can avoid interference with the real-time system on the premise of supporting security detection. However, building and maintaining this platform takes a lot of time and money. At this time, designers can use digital twins technology to solve this problem. The digital twins can provide a continuous monitoring record of safety aspects for the entire life cycle of the CPS \cite{tauber2018enabling}. In the Metaverse era, it is bound to be more vital to conduct real-time system monitoring and detection to prevent the real-time invasion of hackers.
    
4) In the work proposed by Damjanovic \textit{et al.} \cite{damjanovic2018digital}, automatic privacy assessment based on virtual replicas of smart cars is studied. The digital twins then anonymize the data before transmitting it to the insurance company, protecting the privacy of the customer. In this way, the technologies of digital twins further strengthens the protection of personal information security, and they can be used for reference in the Metaverse.
    
5) The digital twins provide a platform that can virtualize the process of processing, virtualize the cost of media sanitization, data confidentiality, and clean-up processes, and suggest appropriate solutions for secure decommissioning of CPSs \cite{grieves2017digital}. There are more large-scale systems in the Metaverse era, and a large Metaverse can be built to safely retire the systems. Hence, it has to dispose of the components by referring to the related technologies of digital twins.

In general, since the digital twins technology and many parts of the Metaverse have similar concepts, the security technology in the digital twins can provide a powerful reference for the future security issue of the Metaverse.

\section{Open Problems and Opportunities}
\label{sec:opportunities}

\subsection{Open Problems}

Although existing technology has brought many solutions to the problem of security and privacy in the Metaverse. However, it is undeniable that there are still some open problems that remain unresolved.

\subsubsection{NFTs (Non-Fungible Tokens) \upshape{\cite{wang2021non}}} The Metaverse will rely on NFTs. In other words, NFTs are the foundation of the Metaverse. There are integrity issues. When users use NFTs for transactions, it can be understood that the transaction data is used as an asset receipt. However, the assets of these receipts exist on the server, and the owner of the server that stores the NFTs is the actual controller of the transaction data. From another point of view, once these centralized exchanges determine that the NFTs released by the project have violated regulations for various reasons, causing the work to be removed from the exchange, the NFT consumers' own will become worthless, like the example of the \textit{Not Okay Bears}\footnote{https://notokay.art/} being taken down by \textit{Opensea}\footnote{https://nftevening.com/not-okay-bears-nft-collection-delisted-from-opensea/}.

\subsubsection{Darkverse} The darkverse in Metaverse is like the dark web in the web. In some ways, it is potentially more dangerous than the dark web because of the pseudo-physical presence of the users. The darkverse has no index. Hence, it is more difficult to control.

\subsubsection{Privacy issues} There are still some uncontrollable threats to personal privacy issues. A Metaverse publisher will control all aspects of its Metaverse and collect a large amount of user data. Due to the decentralized nature of the Metaverse, the Metaverse publisher can monetize and sell this collected data, an act that is also difficult to control.

\subsubsection{Social engineering} Social engineering is able to use psychological manipulation to trick users into creating security breaches or giving away sensitive information. In the Metaverse, criminal disguises will be easier and more difficult to detect. Criminals can infiltrate the Metaverse to impersonate companies, providers, family members, friends, etc. Furthermore, law enforcement agencies may struggle to intercept crimes and criminals in the Metaverse.

\subsubsection{Miscellaneous threats and issues} There are some other security threats to the Metaverse but in various parts, they do not quite fit the categories of the previous issues. These threats and issues include:

\begin{itemize}
    \item The Metaverse has an impact on the environment. For instance, bitcoin mining plays an important role in the Metaverse, which undoubtedly consumes a lot of electricity.
    
    \item The Metaverse has become more heavily reliant on networks. Hence, network malfunctions due to up-link or power failures need to be handled securely.
    
    \item Large companies often have high control over computing power, algorithms, and so on. Therefore, the Metaverse can hardly be disassociated from large technology companies. In other words, it is difficult to achieve absolute decentralization.
    
    \item When decentralization in the Metaverse is fully realized, free from the influence and constraints of the political system and state, it may lead to terrorism running rampant in the Metaverse.
    
    \item In the Metaverse, there will be more infringements due to anonymity. Policies and enforcement of copyright need to be refined and supplemented.
    
    \item More technologies related to artificial intelligence, robotics, etc. will appear, but there is still a lack of ethics, accountability, and laws governing interactions with bots or artificial intelligence.
    
    \item The speech and activities within the Metaverse may be flooded (fake news, hate, extremism, racism, bullying, harassment, and so on).
\end{itemize}

\subsection{Future Directions}

For a variety of reasons, security and privacy remain the most pressing concerns about the Metaverse today. With the coming of a new age of digital information exploration, it will become a critical area to structure the virtual world with the tools to solve the issues of security and privacy. Therefore, we still have a lot of work to construct a blueprint and further solve how the virtual world integrates with the real world.

\subsubsection{Vigorously developing science and technology} We have mentioned in Section \ref{sec:security} that many existing technologies have supported the security of the Metaverse. With the emergence of unknown areas, there will inevitably be higher technical requirements. For example, developing computing power, developing new algorithms to improve computing speed, opening up new storage technologies, and realizing further innovation on the basis of existing interactive technologies. Enterprises and companies need to be good at healthy competition and cross-border cooperation to ensure that the speed of scientific and technological upgrading can maintain the development of the Metaverse.

\subsubsection{Improve laws and regulations} Since the network was born, it has developed so fast that many applications are widely used in a variety of fields. However, it didn't take long for security concerns to arise, like robots, inappropriate speech, national questions, and so on. Actually, due to the Metaverse, many issues have to be rethought again. Many things conceived twenty years ago must be reconsidered and updated before the Metaverse can be fully constructed. The order of the Metaverse will undoubtedly be more difficult to manipulate and govern due to anonymity and decentralization. The future regulation of law in the Metaverse will be a significant challenge. For example, the legislature must define rights, create accountability, protect responsibilities for data, and reinforce the standard of authentication in the Metaverse. In the future, laws and regulations will need to be improved to maintain the order of the Metaverse and provide a fundamentally secure virtual world.

\subsubsection{Find a balance between regulatory and user experience} For the guarantee of the security and privacy of the Metaverse, government departments may impose excessive intervention and regulation. Although security and privacy issues should not be ignored, excessive supervision may lead to a decline in users' experiences. Hence, the relationship between regulatory and user experiences should be well-balanced. We should commit to a "responsible Metaverse" as much as possible, not only to manage a series of risks and complexities but also to ensure we maintain the trust of users in the Metaverse.

\section{Conclusions}  \label{sec:conclusions}
 
As an emerging topic in the 2020s, the Metaverse provides a lot of possibilities for the future direction of the new generation of the Internet. However, the issue of security and privacy is enduring and cannot be ignored in the Metaverse. In this survey, we comprehensively review the advantages of the Metaverse, the changes it brings to mankind, the application scenarios, and related technologies that help the Metaverse develop. Furthermore, we expound on the security and privacy guarantees brought about by the Metaverse's development through five technologies. In addition, some vital open problems and opportunities are discussed in detail. Finally, we conclude this paper. We hope this survey can make the Metaverse more attractive to both industry and academia.

\section*{Acknowledgment}

This research was supported in part by the National Natural Science Foundation of China (Nos. 62002136 and 62272196), Natural Science Foundation of Guangdong Province (No. 2022A1515011861), Guangzhou Basic and Applied Basic Research Foundation (No. 202102020277), Fundamental Research Funds for the Central Universities of Jinan University (No. 21622416), and the Young Scholar Program of Pazhou Lab (No. PZL2021KF0023). Dr. Wensheng Gan is the corresponding author of this paper.

\bibliographystyle{IEEEtran}
\bibliography{references}

\end{document}